\documentclass[a4paper]{IEEEtran}

\usepackage{amssymb,amsfonts,amstext}
\usepackage{verbatim}
\usepackage{amsmath}
\usepackage{algorithm}
\usepackage{algorithmic}
\usepackage{latexsym}
\usepackage{multirow}

\usepackage[dvips]{graphicx}

\newtheorem{proposition}{Proposition} 
\newtheorem{theorem}{Theorem}

\newtheorem{definition}{Definition}

\newcommand{\bA}{\mathbb{A}}
\newcommand{\bF}{\mathbb{F}}

\newcommand{\Sh}{\mathsf{Sh}}
\newcommand{\Rc}{\mathsf{Rc}}

\newcommand{\be}{\begin{equation}}
\newcommand{\ee}{\end{equation}}

\def\Label#1{\label{#1}\ [\ \text{#1}\ ]\ }
\def\Label{\label}

\begin{document}
\title{Universal Construction of Cheater-Identifiable Secret Sharing Against Rushing Cheaters 
Based on Message Authentication}

\author{
 \IEEEauthorblockN{Masahito Hayashi$^{a,b}$
and Takeshi Koshiba$^c$}\\
  \IEEEauthorblockA{$~^{a}$Graduate School of Mathematics, Nagoya University \\
$^b$Centre for Quantum Technologies, National University of Singapore \\
$~^{c}$Faculty of Education and Integrated Arts and Sciences, Waseda University\\
    Email: {masahito@math.nagoya-u.ac.jp \& tkoshiba@waseda.jp} }
} 

\maketitle

\begin{abstract}
For conventional secret sharing, if cheaters can submit possibly forged shares after observing shares of the honest users in the reconstruction phase, 
they can disturb the protocol and only they can reconstruct the true secret.
To overcome the problem,
secret sharing schemes with properties of the
cheater-identification have been proposed.
Existing protocols for cheater-identifiable secret sharing have
assumed non-rushing cheaters or honest majority.
In this paper, using message authentication, 
we remove both conditions simultaneously, and 
give its universal construction from any secret sharing scheme.
To resolve this end, we explicitly propose the concepts of ``individual identification'' and ``agreed identification''.
For both settings, we provide protocols for cheater-identifiable secret 
sharing. In our protocols, the security parameter can be set 
independently to the share size and the underlying finite field size.
\end{abstract}


\section{Introduction}
Secret sharing is a basic primitive for 
secure information transmission \cite{Shamir}.
It involves a dealer who has a secret $S$ in the secret space ${\cal S}$ and a set of players.
The dealer divides the secret $S$ into $n$ shares and distributes the shares to $n$ players such that 
if a set of players is qualified then
all the players in the set can reconstruct the secret 
and if the set of players is not 
qualified then any player in the set cannot obtain any information 
about the secret.
In case of $(k,n)$-threshold scheme, any set of $k$ players can be
qualified.
Generally, a family $\bA$ of subsets of $\{1, \ldots, n\}$, denoted by $[n]$,
is the access structure of a secret sharing protocol
when all the players in each subsets in $\bA$ can reconstruct the secret $S$ 
and others cannot learn anything about it.
It is known that when a family $\bA$ of subsets is closed 
with respect to the union operation,
there exists a secret sharing protocol whose access structure is $\bA$ \cite{ISN,BSV,IS,KW93,Brickell89e,Brickell89c}.
Further, when a non-qualified set of players obtains a part of information,
these schemes are called non-perfect. 
If a non-perfect scheme has certain threshold properties, 
then it is called a ramp scheme \cite{Blakley,Y85,Farrs}.

\begin{table*} 
\caption{Comparison of proposed CISS protocol for individual identification with existing CISS protocols}
\label{hikaku}
\begin{center}
\begin{tabular}{|c|c|c|c|c|c|c|}
\hline
& Number of  & \multirow{2}{*}{Rushing}  & Universal & \multirow{2}{*}{Efficiency} & \multirow{2}{*}{Flexibility} & Large  \\
& Cheaters &                         & Construction  &   &  & Finite Field \\
\hline
\cite{Rabin} & $t<n'/2=n/2$ & No   & No  & $O(\ell \log \ell)$ & Yes & Needless \\
\hline
\cite{IOS12} & $t<n'=n$ & No   & Yes  & $O(\ell \log \ell)$ & No & Needless \\
\hline
\cite{RAX,XMT1,XMT2} & $t<n'/2=k/2$ & Yes  & No  & $O(\ell \log \ell)$ & No & Need \\
\hline
\cite{PW91} & $t<n'/2=k/2$ & Yes  & No  & $O(\ell \log \ell)$ & Yes & Need \\
\hline
Proposed & $t<n'$ & Yes   & Yes  & $O(\ell \log \ell)$ & Yes & Needless\\
\hline
\end{tabular}
\end{center}
$n$ is the number of the players.
$n'$ is the number of the applicants.
$t$ is the number of the cheaters.
$k$ is the number of qualified players. 
$1-2^{-\ell}$ is the successful probability to identify the cheaters.
Efficiency shows the computational complexity of the protocol.
Flexibility is the independence of the choice of the security parameter $\ell$ from 
the secret size or the form of original protocol.
\end{table*}

For conventional secret sharing protocols, 
it is assumed that everyone involved in the protocols is honest or semi-honest. 
However, in a real scenario, some applicants may maliciously behave in the execution of the protocol. 
In particular, a part of players may submit incorrect shares so 
as to yield an incorrect secret in the reconstruction phase. 
To overcome the problem, additional properties to conventional secret sharing 
have been considered and new schemes such as
cheater-detectable secret sharing (CDSS) \cite{TW} and cheater-identifiable secret sharing (CISS) \cite{MS}
have been proposed.
Here, a player that submits incorrect shares is called a cheater.
A protocol is called a $(t,\epsilon)$-cheater-detectable secret sharing (CDSS)
when it detects the existence of cheaters among players involved in the reconstruction phase with probability $1-\epsilon$ at least
under the condition that the number of cheaters is not greater than $t$.
A protocol is called a $(t,\epsilon)$-cheater-identifiable secret sharing (CISS)
when it identifies who submitted incorrect shares with probability $1-\epsilon$
at least
under the condition that the number of cheaters is not greater than $t$.

Even in the setting of CISS,
cheaters may submit their shares incorrectly {\it after} observing shares of honest players. 
Such cheaters is called {\it rushing} cheaters.
The papers \cite{RAX,XMT1,XMT2,PW91} proposed CISS protocols to properly works against such rushing cheaters.
To achieve this task, their sharing phase is composed of two rounds.
Unfortunately, these protocols cannot identify the cheaters
when the number of cheaters is more than half of players involved in the
reconstruction phase.
In this situation, only the protocol in \cite{PW91} can detect the existence of cheaters without identifying them.
Ishai et al \cite{IOS12} proposed another CISS protocol identifying them even when the number of cheaters is more than half of players involved in 
the reconstruction phase.
To achieve this task, they propose a locally-identifiable secret sharing (LISS), in which 
a server identifies the cheaters instead of each player,
but their LISS is not robust against rushing cheaters.
In their protocol, the players submit their shares to the server,
and the server recovers the secret and identifies the cheaters for each player.
While the server sends each player an information to identify the cheaters,
this information depends on the player.
That is, this information is correct only when the player is honest.
Hence, their identifications do not agree in this protocol.

In a real scenario, it is not easy to prepare the server.
Therefore, it is strongly required to propose a protocol to identify the rushing cheaters
even when more than half of the players involved in the reconstruction phase are cheaters.
In this paper, to resolve this problem,
we explicitly propose the concepts of ``individual identification'' and ``agreed identification''.
A CISS protocol with {\it individual identification} privately identifies the cheaters so that the identification depends on individual players. 
A CISS protocol with {\it agreed identification} commonly identifies the cheaters so that the identification is independent of the player. 
The difference between these two types of protocols is based on 
whether their identifications agree or not. 
The protocol in \cite{IOS12} belongs to the former, and  
the protocols in \cite{RAX,XMT1,XMT2,PW91} do to the latter.
In the case of CDSS protocols,
we do not have to care about the difference
because it is not advantageous for CDSS protocols to individually detect 
the existence of the cheaters.

In this paper,
we propose a CISS protocol with individual identification 
as well as a CISS protocol with agreed identification.
Both protocols well work even against rushing cheaters, and
the latter is composed of two rounds as well as the protocol in \cite{PW91}.
The former can identify the cheaters even when 
more than half of the players
involved in the reconstruction phase are cheaters.
The latter can detect the existence of the cheaters under the same situation, 
but can 
identify the cheaters only when 
less than half of the players in the reconstruction phase are cheaters.
When less than half of the players involved in the reconstruction phase are cheaters,
even the latter can identify the cheaters.
This performance is the same as the protocol given in \cite{PW91}.

Next, we discuss the construction of protocols.
Algebraic structures underlie many CISS protocols
\cite{RAX,XMT1,XMT2,PW91}
as in the original construction by Shamir.
They are limited to $(k,n)$-threshold scheme protocols.
However, 
so many efficient secret sharing protocols were proposed when the size of secret is large \cite{Y85,IS,Blakley,Chen}.

Protocols with general access structures were constructed \cite{ISN,BSV,IS}.
Also, ramp-type secret sharing protocols were constructed \cite{Blakley,Y85}.
Such general secret sharing protocols were not used to in these CISS protocols.
Hence, it is desired to construct a CISS protocol by converting an existing secret sharing protocol.
Such a construction is called a {\it universal} construction.
The protocol in \cite{IOS12} is universal in this sense. 
But, it was constructed by converting an existing secret sharing protocol 
only when the share is given as an element of a finite field.
So, to make the scheme more secure,
it needs a finite field of larger size. 

For a precise analysis, we need to distinguish a player wishing to open the secret so called an applicant
from a qualified player because the set of applicants contains cheaters,
while some of existing studies  \cite{PW91,RAX,XMT1,XMT2} assume that 
they are identical.
Hence, the protocol needs to work well when 
the set of applicants is larger than the set of qualified players.
To satisfy this requirement, we characterize qualified players by general access structure,
and universally construct our protocols from an existing arbitrary secret sharing protocol with general access structure
when the share is given as an element of vector space of a finite field.
Our universal construction of CISS for both settings employs
the method of the message authentication based on universal-2 hash function \cite{K95,M96}.
That is, we attach the message authentication \cite{K95,M96} to the share of the existing secret sharing protocol.
Then, the identification can be realized by checking whether the share is original. 
Hence, our construction does not require a finite field of large size.

From a practical viewpoint, we need to care about the computational complexity of the protocol.
A protocol is efficient when its computational complexity is not so large.
When the players identify the cheaters with probability $1-2^{-\ell}$ where
$\ell$ is the security parameter,
the computational complexity of the protocols given in \cite{PW91} is $O(\ell \log \ell)$.
When the protocol is universally constructed,
the total computational complexity depends on the original secret sharing protocol.
In this case, we focus on computational complexity except for the part of the original protocol.
In this sense, the computational complexity of
the protocol in \cite{IOS12} is $O(\ell \log \ell)$, and that of our protocol is also $O(\ell \log \ell)$.

In general, the security parameter $\ell$ may depend on the
other parameters for the protocols.
In the protocols in \cite{RAX,XMT1,XMT2},
the security parameter $\ell$ depends on the size of secret.
On the other hand, it is desirable 
to flexibly choose the security parameter $\ell$.
We call a protocol flexible, when the security parameter $\ell$ can be set independently, i.e., independent of the secret size. 
Flexibility provides the power of partial customization of length of random strings, according to the requirement.
The protocol in \cite{PW91} can flexibly choose the security parameter $\ell$
by adjusting the finite field with prime size. 
Also, the protocol in \cite{IOS12} can flexibly choose the security parameter $\ell$
by adjusting the finite field appearing in the original protocol.
Although these protocols offer the flexibility, 
the security parameter $\ell$ depends on the size of the finite field.
The above computational complexity $O(\ell \log \ell)$ can be realized 
by suitable choices of the size of the finite field in these protocols \cite{Hayashi-Tsurumaru}.
Hence, the choice of the security parameter $\ell$ has a certain restriction 
when we keep the computational complexity $O(\ell \log \ell)$.
Therefore, it is desirable to completely freely choose the security parameter $\ell$.
Fortunately, our protocol works with any finite field, and 
the security parameter $\ell$ can be freely chosen independently of the size of the finite field and the secret size. 
Therefore, our protocol is completely flexible and works even with finite field $\bF_2$, which simplifies the realization.
Overall, the comparison of the performances of existing protocols with ours 
is summarized as Table \ref{hikaku}.

The remaining part of this paper is as follows.
Section \ref{sec:prep} gives basic tools and formal definitions.
Section \ref{S2} gives our CISS protocol for individual identification.
Section \ref{S3} shows its security.
Section \ref{S4} gives our CISS protocols for agreed identification and detection.
Section \ref{S5} compares the overhead of ours with those of existing protocols.

\begin{figure*}
\begin{center}
\scalebox{1}{\includegraphics[scale=0.48]{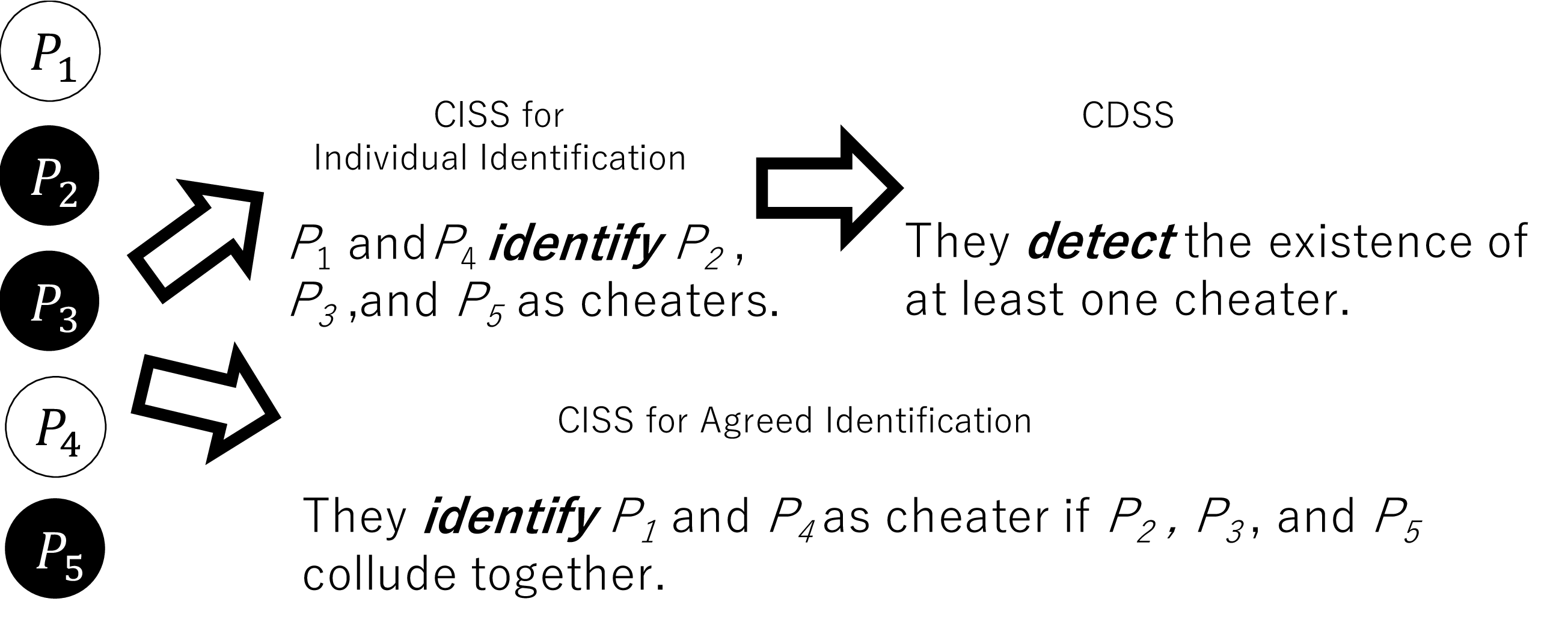}}
\end{center}
\caption{A Case of majority cheaters.
A white circle expresses a honest player and
black circles express cheaters.}
\Label{F1}
\end{figure*}%

\section{Preliminaries}\label{sec:prep}
\subsection{Notations}
If $A$ is a finite set, $a\leftarrow A$ means that an element $a$ is uniformly
chosen from $A$. If $A$ is a probability distribution, $a\leftarrow A$
means that $a$ is chosen according to $A$. If $A$ is a probabilistic algorithm,
$A(\cdot)$ can be regarded as a probability distribution. 
Thus, $a\leftarrow A(\cdot)$ can be defined as in the case of probability
distributions. For any event $\bf E$ depending on a random variable $A$, 
$\Pr_{a\leftarrow A}[{\bf E}]$ means that $\sum_{a\in {\rm supp}(A)} \Pr[A=a]\cdot\Pr[{\bf E}\mid A=a]$.

\subsection{Universal Hash Functions}
As mentioned, our universal construction uses message authentication protocols
\cite{K95,M96} in which universal hash functions are used. 
A family ${\cal H}$ of hash functions $h_i : A \rightarrow B$ 
is said to be universal-2
if for any $a\in A$ and $b\in B$
\[ \Pr_{h_i\leftarrow \cal H}[h_i(a)=b] = \frac{1}{|B|}
\]
holds, where the probability is over the uniformly random choice of $h_i$
from $\cal H$. The original definition of universal-2 hash functions and
several constructions can be found in \cite{CW79}.

\subsection{Toeplitz Matrices}
An $(n\times m)$-Toeplitz matrix $T=(T_{i,j})$ over $\bF_q$ can be determined by
$s=s_1s_2\cdots s_{n+m-1}\in \bF_q^{n+m-1}\setminus\{0^{n+m-1}\}$ as follows:
Set $T_{n,1}=s_1, \ldots, T_{1,1}=s_n$ (for the first column) and
$T_{1,2}=s_{n+1}, \ldots, T_{1,m}=s_{n+m-1}$ (for the first row).
The remaining entries can be set so as to satisfy
that $T_{i,j}=T_{i-1,j-1}$ for any $i$ and $j$ with $2\le i\le n$ and
$2\le j\le m$. That is, $T$ can be written as
\[ T= \left( \begin{array}{ccccc}
s_n & s_{n+1} & s_{n+2} & \cdots & s_{n+m-1}\\
s_{n-1} & s_n & s_{n+1} & \cdots & s_{n+m-2}\\
s_{n-2} & s_{n-1} & s_n & \cdots & s_{n+m-3}\\
\vdots & \vdots  & \vdots & \ddots & \vdots\\ 
s_1 & s_2 & s_3 & \cdots & s_m
\end{array}\right).
\]  
The matrix $T$ represents a linear map from $\bF_q^m$ to $\bF_q^n$.
If $n<m$, $T$ can be regarded as a hash function indexed by $s$.
It is known that Toeplitz matrices work as universal-2 hash functions
with short description. 

\begin{proposition}
Let ${\cal H}_{m,n} = \{ (m\times n)$-Toeplitz matrix $T$ determined by $s\mid
s\in \bF_q^{n+m-1}\setminus\{0^{n+m-1}\}\}$, where $n<m$. Then ${\cal H}_{m,n}$ is a family of
universal-2 hash functions from $\bF_q^m\setminus\{0^m\}$ to $\bF_q^n$.
\end{proposition}

We would like to remark that universal-2 hash functions
based on Toeplitz matrices are advantageous. In general, universal-2 hash
functions require the uniform distribution over the hash function class 
$\cal H$. Even if there is a bias in the distribution over $\cal H$ in the
sense of \cite{NN93}, the resulting hash functions are still ``almost''
universal \cite{K95}. This property of the robustness is applicable to
our results in this paper, although we do not discuss this matter any more.

\subsection{Secret Sharing}
A secret sharing protocol $(\Sh,\Rc)$ consists of
two subprotocols $\Sh$ and $\Rc$.
The dealer, having a secret $s$ in the secret space $\cal S$,
initiates the sharing protocol $\Sh$. At the end of $\Sh$, 
the $i$-th player for each $i\in [n]$ obtains
his share $v_i$ in the share space $\cal V$. In the standard secret sharing
schemes, $\Sh$ is just a non-interactive protocol (say, a probabilistic 
algorithm), which takes $s\in\cal S$ as input and produces
$v_1,\ldots,v_n$. After producing $n$ shares, the dealer sends
$v_i$ to the $i$-th player for each $i\in [n]$.
We may regard $\Sh$ as a (probabilistic) function from $\cal S$ to ${\cal V}^n$.

Usually, $\Rc$ is defined as an algorithm which takes
a subset of all the shares with players' indices as input and 
tries to recover the secret $s$.
In the real situation, if a player wishes to recover the secret, he
should declare his wish to do that and collet shares from the other players
that also wish to recover the secret. We call such players {\em applicants}.
If the set of all the applicants is so-called qualified, 
the reconstruction algorithm 
should recover the secret $s$. Otherwise, the secret should not be recovered.
The qualification can be defined in terms of access structures.
An access structure $\bA$ is a monotone collection of sets of players.
Any set $A$ is said to be qualified if and only if $A\in \bA$. Moreover,
if $A'\supset A$ for a qualified set $A$, $A'$ should be also qualified.
In this paper, we do not give definitions of the correctness of the 
reconstruction algorithm. This is because our results are transformations
from a underlying secret sharing scheme to another scheme with the
cheater identification. The resulting scheme inherits 
the correctness property and the access structure from the underlying scheme. 
For example, if the underlying scheme has the perfect correctness then
the resulting scheme also has the perfect correctness, and if the
underlying scheme is a ramp scheme then the resulting scheme is also a
ramp scheme.

As mentioned, the applicants declare their wish to recover the secret.
Before executing the reconstruction algorithm, the applicants must collect
shares to feed the algorithm with by invoking some {\em preparation} protocol. 
Normally in the preparation phase, applicants may publish their shares or send
them to each other. In our universal constructions, we will explicitly
mention the preparation phase.


\subsection{Cheater Identifiability and Detectability}
Let $\cal P$ be a set of players and $\cal P$ can be usually identified
with $[n]$. 
Reconstruction algorithms for the standard secret sharing schemes
can be regarded as a map from $2^{{\cal P}\times {\cal V}}$ to 
${\cal S} \cup \{\bot\}$ where
$\bot$ denotes a failure of the reconstruction.
On the other hand,
reconstruction algorithms for cheater-identifiable schemes
are regarded as a map from $2^{{\cal P}\times {\cal V}}$ to $({\cal S} \cup \{\bot\})
\times 2^{\cal P}$. That is, $\Rc$ outputs $(s',L)$ where $s'\in {\cal S}\cup \{\bot\}$
and $L$ is a subset (or list) of applicants. We expect that $s'=s$ and $L$ is 
the list of all cheating applicants.

To give formal definitions of the cheater identifiability, we consider
the following experiment ${\bf Exp}_{\cal A}$ for the adversary 
${\cal A}=({\cal A}_1,{\cal A}_2)$.
\begin{quote}
\begin{tabular}{ll}
1.&$s\leftarrow {\cal S}$;\\
2.&$(v_1,...,v_n)\leftarrow \Sh(s)$;\\
3.&$B\leftarrow$ the set of honest applicants;\\
  &{\tt\verb+/*+} Honest applicants declare their wish\\
  &\hspace*{1cm} to recover the secret.
   {\tt\verb+*/+}\\
4.&$(C,{\it state})\leftarrow {\cal A}_1(B)$;\\
&{\tt\verb+/*+} $C$ is the set of applicants corrupted by $\cal A$\\ 
&\hspace*{1cm}satisfying $B\cap C=\varnothing$ {\tt\verb+*/+}\\
5.& Honest applicants send their shares $\{v_i\}_{i\in B}$\\
& to the other applicants (including corrupted ones);\\
6.&Corrupted applicants forge their shares by\\
& computing $\{v_{i,j}'\}_{i\in C,j\in B}\leftarrow {\cal A}_2(\{v_i\}_{i\in B\cup C},{\it state})$\\
& and send $\{v_{i,j}'\}_{i\in C}$ to the $j$-th honest applicant\\
& for each $j\in B$;\\
7.& Honest applicants individually execute the\\
& reconstruction algorithm as
$\{Out_j\}_{j\in B}\leftarrow$\\
&$\{\Rc(\{i,u_{i,j}\}_{i\in B\cup C})\}_{j\in B}$, where $u_{i,j}=v_i$\\
& if $i\in B$ and $u_{i,j}=v_{i,j}'$ otherwise.
\end{tabular}
\end{quote}
In the above, the output $Out_j$ of the reconstruction algorithm
executed by the $j$-th applicant is of the form $(s',L_{j,\cal A})$.
Let ${\bf E}_{i,j,{\cal A}}$ for each $i\in B\cup C$ and $j\in B$ 
be the event that $i$ is not included in the list $L_{j,\cal A}$ 
The event ${\bf E}_{i,j,{\cal A}}$ 
means that the $j$-the applicant does not consider the
$i$-th applicant corrupted by the adversary $\cal A$.

\begin{definition}\label{def:ciii}
A secret sharing scheme $(\Sh,\Rc)$ is $(t,\varepsilon)$-
cheater-identifiable for individual identification if the following
condition holds. For any adversary $\cal A$ which collapses at most $t$
players and for any two distinct applicants $i,j$,
\begin{itemize}
\item $\Pr[{\bf E}_{i,j,\cal A}]\le \varepsilon$ if 
the $j$-th applicant is honest and 
the $i$-th applicant submits a forged share $v_{i,j}'\ne v_i$ to 
the $j$-th applicant.
\item $\Pr[{\bf E}_{i,j,\cal A}] = 1$ if 
the $j$-th applicant is honest and
the $i$-th applicant
submits the original share $v_{i,j}'= v_i$ to the $j$-th applicant.
\end{itemize}
\end{definition}

\begin{definition}\label{def:ciai}
A secret sharing scheme $(\Sh,\Rc)$ is $(t,\varepsilon)$-
cheater-identifiable for agreed identification if 
for every adversary $\cal A$ who collapses at most $t$ players and 
always forges shares and 
for any two distinct applicants $i,j$, the same conditions as
in Definition \ref{def:ciii} hold and
all lists $L_{j,\cal A}$ for honest applicants $j$ coincide with each other.
\end{definition}

\begin{definition}\label{def:cd}
A secret sharing scheme $(\Sh,\Rc)$ is $(t,\varepsilon)$-
cheater-detectable if for every adversary $\cal A$ who collapses 
at most $t$ parties, there exists a honest applicant $j$ such that
$\Pr[L_{j,\cal A}=\varnothing]\le \varepsilon$ if the $j$-th applicant
receives forged shares.
%
\end{definition}


\section{Protocol for Individual Identification}\Label{S2}
Let $n$ be the number of players and $\ell'$ be the security parameter, which 
is independent of the structure of the original secret sharing protocol.
We will construct our protocol so that the verifier (honest applicant) 
identifies the cheater with probability more than $1-q^{-\ell'}$.
While our construction is given by attaching the message authentication protocol \cite{K95,M96} to the share of an existing secret sharing protocol like \cite{RAX},
it works with an arbitrary existing secret sharing protocol.

Let $(\Sh,\Rc) $ be a secret sharing protocol realizing access structure $\bA$
with $\Sh :{\cal S} \to {\cal V}^n$, where $\cal V$ is $m$-dimensional vector space $\bF_q^m$ over a finite filed $\bF_q$.
Here, the access structure $\bA$ expresses the condition for a subset of $[n]$
to recover the secret $S\in {\cal S}$.
To present our CISS protocol for individual identification based on the protocol $(\Sh,\Rc) $,
we make preparation as follows.
We employ message authentication based on Toeplitz matrices,
which are often used in cryptography (e.g., hard-core functions \cite{GL89}
and universal hash functions \cite{K95,MNT90}).
For a secret $S\in {\cal S}$, we define the random number $X_i:=\Sh_i(S)$ 
where $\Sh_i(S)$ is the projection of $\Sh(S)$ onto the $i$-th coordinate,
as 
the share of the $i$-th player, which is sent by the dealer.
For $i \neq j$,
the dealer independently generates $n(n-1)$ random numbers $Z_{j,i}$ taking values in $\bF_q^{\ell'}$.
Also, the dealer independently generates $\ell'\times m$ Toeplitz matrix $T_j$.
Then, the dealer calculates the random number $Y_{j,i}:=  T_j X_i+Z_{j,i}$.
Now, we give our CISS protocol for individual identification as Protocol \ref{protocol1}.
Due to the construction of Protocol \ref{protocol1}, we find that 
its computational complexity is $O(\ell'\log \ell')$.

\begin{Protocol}                  
\caption{CISS protocol for individual identification}         
\label{protocol1}      
\begin{algorithmic}
\STEPONE[Dealing] For each $j\in [n]$,
the dealer sends the $j$-th player 
the publishable information
$(X_j, Z_{1,j}, \ldots,$\break
$Z_{j-1,j},Z_{j+1,j} \ldots, Z_{n,j})$
and the identification-information 
$(T_j, Y_{j,1}, \ldots,Y_{j,j-1},Y_{j,j+1}, \ldots, Y_{j,n})$.

\STEPTWO[Sharing]  
The applicants send their publishable information as follows.
The $i$-th player sends $(X_i, Z_{j,i})$ to the $j$-th player.

\STEPTHREE[Identification]
The $j$-th player checks whether the relation  
\begin{align}
Y_{j,i}=  T_j X_i'+Z_{j,i}' 
\Label{9-29-1}
\end{align}
holds, where the information received from the $i$-th player is $(X_i', Z_{j,i}')$,
which is the same process as the verification of the message authentication \cite{K95,M96}. 

\STEPFOUR[Reconstruction]  
If the set of the players verified by the $j$-th player to be honest satisfies the access structure $\bA$,
the $j$-th player reconstructs the secret from the collection of $X_i'$ of players verified to be honest.
\end{algorithmic}
\end{Protocol}

\section{Security Analysis}\Label{S3}
As our security analysis, we show the following theorem.

\begin{theorem}
Let $n'$ be the number of applicants.
Protocol \ref{protocol1}
is an $(n'-1,q^{-\ell'})$-CISS protocol for individual identification realizing access structure $\bA$ with secret space ${\cal S}$
and share space ${\cal S}_i=\bF_q^{(2 n -1)\ell'+2 m-1}$.
Further, if the set of honest players satisfies the access structure $\bA$.
each player can reconstruct the secret $S\in {\cal S}$
\end{theorem}

\begin{IEEEproof}
Since the function $(X_i, Z_{j,i}  ) \mapsto T_j X_i+Z_{j,i}$
is a universal-$2$ hash function with the randomly chosen Toeplitz matrix $T_j$,
the relation \eqref{9-29-1} holds with probability smaller than $q^{-\ell'}$
if the $j$-th player makes a cheat.
Therefore, even though all of players except for the $i$-th player makes cheating
even with collusion,
the $i$-th player can identify who makes cheating with high probability as Fig. \ref{F1}.

Also, even though several players collude together,
they cannot obtain any information for the shares by other players as follows.
To see this fact,
we assume that the $j_1$-th player, the $j_2$-th player, ...,
the $j_a$-th player 
collude together.
We focus on the information on $X_i$ shared by the $i$-th player.
Since $Z_{j,i}$ is independent and uniform,
$Y_{j_1,i}, Y_{j_2,i}, \ldots, Y_{j_a,i}$ are independent of $T_{j_1} X_i, T_{j_1} X_i,\ldots, T_{j_a} X_i$.
Since they obtain no information for  
$T_{j_1} X_i, T_{j_1} X_i,\ldots, T_{j_a} X_i$, they obtain no information on $X_i$.

Thus, if the original protocol with shares $X_i$ works as secret sharing well,
our protocol also works as secret sharing well 
at least with probability $1-q^{-\ell'}$.
Therefore, we obtain the desired statement.
\end{IEEEproof}



\section{Protocol for Agreed Identification}\Label{S4}
Now, we can give our CISS protocol for agreed identification as Protocol \ref{protocol2}.

\begin{Protocol}                  
\caption{CISS protocol for agreed identification}         
\label{protocol2}      
\begin{algorithmic}
\STEPONE[Dealing] For each $j\in [n]$, 
the dealer sends the $j$-th player 
the publishable information 
$(X_j, Z_{1,j}, \ldots,$\break
$Z_{j-1,j},Z_{j+1,j} \ldots, Z_{n,j})$
and the identification-information 
$(T_j, Y_{j,1}, \ldots,Y_{j,j-1},Y_{j,j+1}, \ldots, Y_{j,n})$.

\STEPTWO[Sharing (Round 1)]  
The applicants send their
publishable information. 

\STEPTHREE[Sharing (Round 2)]  
The applicants send their
identification-information.

\STEPFOUR[Identification]
Each player applies respective individual identification.
We employ the majority voting of the results of respective individual identification.
That is, if $\lceil (n'-1)/2\rceil$ players agree to identify the set of honest players,
the set is identified to the set of players identified to be honest.

\STEPFIVE[Reconstruction]
If the above set of players identified to be honest satisfies the access structure $\bA$,
the players reconstruct the secret from the collection of $X_i'$ of players verified to be honest.
\end{algorithmic}
\end{Protocol}

Since the majority voting of the results of respective individual verifications 
identifies (as shown in Theorem \ref{protocol1})
who makes cheating if more than half of the players wishing the reconstruction are honest,
we have the following theorem.
\begin{theorem}
Let $n'$ be the number of applicants.
Protocol \ref{protocol2}
is a $( \lceil (n'-1)/2\rceil,q^{-\ell'})$-CISS protocol realizing access structure $\bA$ with secret space ${\cal S}$
and share space ${\cal S}_i=\bF_q^{(2 n -1)\ell'+2 m-1}$.
Further, if the number $t$ of cheaters satisfies $t < \lceil (n'-1)/2\rceil$ 
and the set of honest players satisfies the access structure $\bA$,
the players agreeably reconstruct the secret $S\in {\cal S}$.
\end{theorem}

Further, we can make a CDSS protocol
by modifying Protocol \ref{protocol2} in the following way,

\begin{Protocol}                  
\caption{CDSS protocol}         
\label{protocol3}      
\begin{algorithmic}
\STEPONE[Dealing] The same as Protocol \ref{protocol2}.

\STEPTWO[Sharing (Round 1)]  
The same as Protocol \ref{protocol2}.

\STEPTHREE[Sharing (Round 2)]  
The same as Protocol \ref{protocol2}.

\STEPFOUR[Detection]
Each player applies respective individual identification.
If there exists a player who individually identifies at least one cheater, 
we consider that there exists a cheater.

\STEPFIVE[Reconstruction]
If no cheater is detected,
the players reconstruct the secret from the collection of $X_i'$.

\end{algorithmic}
\end{Protocol}

If there exists a player who individually identifies
at least one cheater, we consider that there exists a cheater.
So, Protocol \ref{protocol2}$'$ detects the existence of the cheaters with probability 
$1-q^{-l'}$ as Fig. \ref{F1}, which yields the following theorem.

\begin{theorem}
Let $n'$ be the number of applicants.
Assume that when a subset of $[n]$ contains $n'$ elements,
it satisfies the access structure $\bA$.
Protocol \ref{protocol3}
is an $( n'-1,q^{-\ell'})$-CDSS protocol realizing access structure $\bA$ with secret space ${\cal S}$
and share space ${\cal S}_i=\bF_q^{(2 n -1)\ell'+2 m-1}$.
Further, if there is no cheater,
the players agreeably reconstruct the secret $S\in {\cal S}$.
\end{theorem}

Now, we consider the case when more than half players collude together.
We assume that only the $j_0$-th player is honest
and that 
the majority cheaters (the $j_1$-th player, ..., the $j_a$-th player) 
collude together.
The cheater, the $j_v$-th player rewrites $T_{j_v}$, $Z_{j_v,j_w}$ and $Y_{j_v,j_w}$ for 
$1 \le v \le a$, $0 \le w \le a$
so that 
$Y_{j_v,j_w}=  T_{j_v} X_{j_w} +Z_{{j_v},j_w} $ for $1 \le w \le a$
and
$Y_{j_v,j_0} \neq   T_{j_v} X_{j_0}+Z_{{j_v},j_0} $.
Due to the majority voting,
the agreed identification is that the honest player, the $j_0$-th player is a cheater.
Therefore,
when the majority make cheating,
the identification of our CISS protocol for agreed identification is incorrect
while 
our CDSS protocol detects the existence of a cheater and 
the identification of our CISS protocol for individual identification is correct,
as Fig. \ref{F1}.

\section{Comparison of Overhead}\Label{S5}
First, we compare the overhead of the protocol in \cite{IOS12} with ours.
Let $u$ be the size of the share of the original secret sharing protocol.
Theorem 4 of \cite{IOS12} guarantees the success probability is $1-2^{-\ell}$ when $2^{-\ell} > n^2(n+1)(u-1)^{-1}$
and 
the size of the share of their CISS protocol 
is $u^{4n+2}$.
That is, their overhead is $u^{4n+1}$.
Since $u> n^2(n+1) 2^{\ell}$, their overhead is 
greater than
$(n^2(n+1))^{4n+1} 2^{\ell (4n+1)}$.
In our case, we have $u= q^m$ and $2^{-\ell} = q^{-\ell'}$.
Hence, our overhead is $ 
2^{\ell (2n-1)}u^{\frac{2m-1}{m}-1}
=2^{\ell (2n-1)}u^{1-\frac{1}{m}}$.
Now, we consider the case with $m=1$ because 
Theorem 4 of \cite{IOS12} considers only this case.
Then, our overhead is $2^{\ell (2n-1)}$.
Therefore, the their exponential coefficient with respect to the security parameter $\ell$ is twice as ours.

Next, we compare the overhead of the protocol in \cite{PW91} with ours.
Since their protocol is specified to the $(k,n)$-threshold scheme,
we translate our overhead to the $(k,n)$-threshold scheme.
When the secret size is $|{\cal S}|$, the conventional $(k,n)$-threshold scheme
has share size $ |{\cal S}| p(n)$
for some polynomial $p$.
When we construct our CISS protocol based on this secret sharing protocol,
the share size is $ |{\cal S}| p(n) 2^{(2 n -1)\ell} q^{m-1}$.
That is, its exponential coefficient with respect to the security parameter $\ell$ 
is still $(2 n -1) $.
In contrast,
the $(\lceil k/2 \rceil, 2^{-\ell})$-CISS protocol in \cite{PW91} 
has share size $ (n-\lceil k/2 \rceil)^{n+k} 2^{(n+k)\ell}$.
See Table 2 of \cite{PW91} with $\epsilon= 2^{-\ell}$
and $t= \lceil k/2 \rceil$.
That is, its exponential coefficient with respect to the security parameter $\ell$ is $ n +k $.
So, when $k$ is close to $n$, 
these two overheads are almost the same.

\section{Discussion}\Label{S6}
Firstly, we have proposed to explicitly distinguish
CISS protocols for individual identification from 
CISS protocols for agreed identification. 
Then, attaching an message authentication protocol to 
any existing secret sharing protocol, 
we have universally constructed
CISS protocols for individual identification and agreed identification 
as well as a CDSS protocol.
Our CISS protocol for individual identification
and our CDSS protocol well work even when more than half
of the players involved in the reconstruction phase are cheaters.
Our CISS protocol for agreed well works when less than half
of the players in the reconstruction phase are cheaters.
Our protocols have computational complexity
$O(\ell \log \ell)$ when the probability of successfully identifying (detecting) the cheaters is $1-2^{-\ell}$. 
We can freely choose the security parameter $\ell$ independently of
the secret size and share size of the original secret sharing protocol.
Also, we do not use huge finite fields.
That is, we can realize any security parameter $\ell$ even with the finite field $\bF_2$.
These characteristics simplify the realization.
We have checked that 
the overhead of our protocols are not so huge in comparison with existing protocols.
Indeed, although our protocol has been given as a simple combination of a message authentication \cite{K95,M96} and an existing secret sharing,
the proposed protocol achieves performances that had not been realized in existing protocols (See Table 1).
In this paper, we have employed message authentication based on Toeplitz matrix for simple performance analysis.
We can generalize our method with a general 
message authentication protocol, which requires more complicated 
performance analysis.

\end{document}